\documentclass[preprint2]{aastex}

\begin{document}

\title{A {\it Chandra} Observation of the Diffuse Emission in the
Face-on Spiral NGC 6946}

\author{Eric M. Schlegel\altaffilmark{1}, S. S. Holt\altaffilmark{2},
R. Petre\altaffilmark{3}}

\altaffiltext{1}{High Energy Astrophysics Division,
Harvard-Smithsonian Center for Astrophysics, Cambridge, MA 02138}

\altaffiltext{2}{Olin College of Engineering, Needham, MA}

\altaffiltext{3}{Laboratory for High Energy Astrophysics, NASA-Goddard
Space Flight Center, Greenbelt, MD}

\begin{abstract} 

This paper describes the {\it Chandra} observation of the diffuse
emission in the face-on spiral NGC 6946.  Overlaid on optical and
H${\alpha}$ images, the diffuse emission follows the spiral structure
of the galaxy.  An overlay on a 6 cm polarized radio intensity map
confirms the phase offset of the polarized emission.  We then extract
and fit the spectrum of the unresolved emission with several spectral
models.  All model fits show a consistent continuum thermal
temperature with a mean value of 0.25$\pm$0.03 keV.  Additional
degrees of freedom are required to obtain a good fit and any of
several models satisfy that need; one model uses a second continuum
component with a temperature of 0.70$\pm$0.10 keV.  An abundance
measure of 3$^{+1.95}_{-1.90}$ for Si differs from the solar value at
the 90\% confidence level; the net diffuse spectrum shows the line
lies above the instrumental Si feature.  For Fe, the abundance measure
of 0.67$\pm$0.13 is significant at 99\%.  Multiple gaussians also
provide a good fit.  Two of the fitted gaussians capture the O~VII and
O~VIII emission; the fitted emission is consistent with an {\it
XMM-Newton} RGS spectrum of diffuse gas in M81.  The ratio of the two
lines is $<$0.6-0.7 and suggests the possibility of non-equilibrium
ionization conditions exist in the ISM of NGC 6946.  An extrapolation
of the point source luminosity distribution shows the diffuse
component is not the sum of unresolved point sources; their
contribution is at most 25\%.

\end{abstract}


\section{Introduction}

The total X-ray emission from a spiral galaxy consists of the sum of
the emission from its X-ray-emitting constituents including X-ray
binaries (low-mass and high-mass), supernova remnants, OB star
clusters, other point sources, and the hot component of the
interstellar medium.  This last component is expected to be present
based upon the theoretical work of \cite{MO77} and \cite{NI89} among
many others.  Prior to the launch of {\it Chandra} and {\it
XMM-Newton}, the hot component had been detected in a handful of
galaxies, largely as a consequence of the broad spatial resolution of
previous X-ray telescopes which distributed X-rays from point sources
over a wide region of the galaxy, contaminating any signal from the
hot ISM.  

The study of the X-ray emission from normal galaxies commenced with
the {\it Einstein} observations of galaxies (for a review, see
\citealt{F89}).  The spatial resolution of {\it Einstein}, $\sim$1$'$
(e.g., \citealt{Har79}), prevented separation of the X-ray emitting
constituents for all but the nearest galaxies.  {\it ROSAT} improved
the resolution by about three times, providing views of a broader
range galaxies outside the Local Group (e.g., \citealt{RobWar01}).
The use of {\it Chandra} and {\it XMM-Newton} has led to more
detections because of the large effective area of {\it XMM-Newton} and
the sharp point spread function of {\it Chandra}.

NGC 6946 is a well-studied face-on spiral initially detected in X-rays
with the {\it Einstein Observatory} \citep{FT87}.  \cite{Sch94}
obtained a 36-ksec pointing using the {\it ROSAT} PSPC, detected 9
point sources, and reported the existence of a diffuse component with
a temperature of 0.55 keV (90\% error range of 0.3-0.65 keV) which we
will show just overlaps with the results to be presented.  That study
argued that the diffuse component was truly diffuse and not the summed
emission of point sources based upon a plausibility argument for the
expected numbers of sources of various types.  In a separate study,
\cite{SBF00} described a {\it ROSAT} HRI observation as well as an
observation using {\it ASCA}.  In this case, the high internal
background of the HRI prevented a detection of the diffuse component,
while the broad PSF of the {\it ASCA} mirrors blended events from the
known point sources with the diffuse emission.

This paper describes the diffuse emission in NGC 6946 revealed by a
60-ksec {\it Chandra} observation of the galaxy.  The point source
population has been described by \cite{Holt02}.  The primary goal of
this paper is a spectrum of the diffuse emission.  Following this
introduction, we describe our analysis steps, then present a brief
comparison of the diffuse emission with images from other bandpasses,
followed by a longer analysis of the spectral fits and a discussion of
the results.

{\it Chandra} presents a unique opportunity to observe the diffuse
component as free as may be possible, at least for the foreseeable
future, from contamination by point sources.  The {\it Chandra}
mirrors deliver $\sim$80\% of the X-rays at 1 keV into a circle
$\sim$0.7 arcsec in radius and 70\% into a 0.7 arcsec radius circle at
6 keV. At 5$'$ off-axis, the corresponding numbers are 90\% into a
circle of a radius $\sim$2$''$.5 at 1 keV and $\sim$4$''$ radius at 6
keV \citep{PropGuide}.

The adopted distance for NGC 6946 is 5.9 Mpc \citep{Kara2000} with a
corresponding scale of 1$''$ $\sim$29 pc.  The column density in the
direction of NGC 6946 is $\sim$3-5$\times$10$^{21}$ cm$^{-2}$ \citep{SFD}.

\section{Observation Summary}

{\it Chandra} observed NGC 6946 on 2001 September 7.  The galaxy was
imaged on the back-illuminated (BI) chip S3.  The data were examined for
flares to which the BI chip is susceptible, but no flares with a count
rate greater than 0.04 counts s$^{-1}$ were located; screening any
flare-like spikes removed 1296 sec of the 59116 sec observation,
leaving a net on-source, deadtime-corrected, exposure of 57819 sec.

The data were filtered to eliminate events with energies below 0.3 keV
and above 8 keV.  The low-energy cut does not remove source photons
because those events are lost to the high column in the direction of
the galaxy.  The high cut removes extraneous cosmic ray-generated
events.

While the internal background of the ACIS detector is low (for S3,
$\sim$0.3 counts s$^{-1}$ chip$^{-1}$, \cite{PropGuide}) and often may
be ignored for counts from point sources extracted with a typical
detection cell a few arc sec in size, the large extent of the galaxy
and the faintness of the diffuse emission require a careful analysis
of the background.  The D$_{\rm 25}$ circle\footnote{The D$_{\rm 25}$
circle is the diameter of the optical isophote at a surface brightness
of 25 mag/arcsec$^2$.} is $\sim$14 arcmin in size \citep{Tul88} and is
larger than the ACIS S3 CCD.  The large extent of the galaxy prevents
the use of portions of the S3 chip outside of the galaxy as a measure
of the background.  We instead obtained the background from a ``blank
field'' observation\footnote{A description as well as links to the
relevant files may be found at
http://cxc.harvard.edu/contrib/maxim/acisbg/.}.  The data were
filtered identically by energy, then re-projected onto the sky using
the aspect solution of the actual NGC 6946 observation.  The exposure
time of the background observation was $\sim$300 ksec.

Point sources were detected as described in \cite{Holt02}, using a
wavelet detection that is directly tied to the point spread function's
behavior with off-axis angle \citep{Free02}.  The point source
detection reached a limiting luminosity L$_{\rm X}$
$\sim$3$\times$10$^{36}$ erg s$^{-1}$ in the 0.3-5 keV band which
corresponded to $\sim$10 counts.  \cite{Holt02} deemed the source
detection complete to $\sim$1$\times$10$^{37}$ erg s$^{-1}$; the
completeness limit is lower by $\sim$30\% near the center of the CCD
because of the sharper point spread function there where $\sim$7-8
counts constitutes a detection.  Most of the detected diffuse emission
lies within 2.5-3 arcmin of the aimpoint.  We do not see a trend of
increasing numbers of weak sources as we approach the aimpoint,
suggesting a relative lack of contamination of the diffuse emission by
weak point sources.  We postpone until \S5 a longer discussion of the
impact of possible unresolved point sources.

\begin{figure*}
\begin{center}
\caption{Chandra observation of NGC 6946 with detected sources
removed.  The data have binned by a factor of 8 to create 4 arcsec pixels.
  ~~ FIGURE IN SEPARATE JPG FILE: 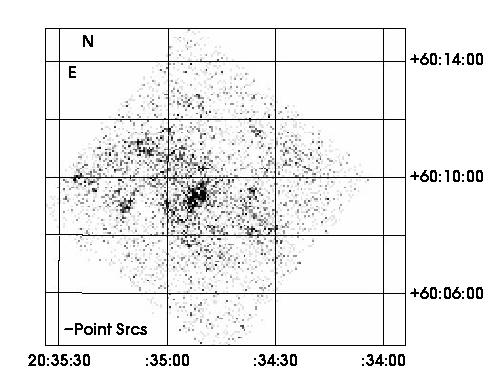 ~~\label{fullobs}}
\end{center}
\end{figure*}

Sources were removed from the data using the CIAO tool {\tt dmfilth}
which cuts out each source, then replaces the resulting `hole' by
sampling the events in annuli surrounding the locations of the
sources.  The extraction apertures used enclosed $>$95\% of the
encircled energy.  The annuli had inner radii 25\% larger than the
radii of the source extraction circles.  Care was taken to ensure that
the annular sampling did not include any source events.
Figure~\ref{fullobs} shows the results of all the steps just
described: the energy-filtered data, minus point sources, and binned
into 4-arcsec pixels.  Immediately from the raw data
(Figure~\ref{fullobs}) we see the diffuse emission as well as bright
areas, particularly on the N and W arms and surrounding the nucleus.

\begin{figure*}
\begin{center}
\caption{Mosaic of the subbands of the diffuse emission, all
adaptively smoothed with a gaussian kernel to place the minimum
background S/N at 3.0.  The 0.3-0.6 0.6-0.9, and 0.9-2.0 keV bands are
shown contours for each band overlaid.  The 0.3-0.6 keV band shows
little emission because of the high column density; contour values are
0.34, 0.38, 45, and 0.55 counts s$^{-1}$ pixel$^{-1}$.  Contours in
the 0.6-0.9 band lie at 0.45, 0.60, 0.75, 0.90, 1.05, and 1.20 counts
s$^{-1}$ pixel$^{-1}$.  Contours in the 0.9-2.0 band lie at 0.50,
0.65, 0.80, 0.95, 1.20, and 1.60 counts s$^{-1}$ pixel$^{-1}$.  The
background levels are 0.26, 0.22, and 0.09 counts s$^{-1}$
pixel$^{-1}$, for the 3 bands, respectively.  The short vertical bar
in the lower right represents a 2 arcmin scale bar.
 ~~~ FIGURE IN SEPARATE JPG FILE:  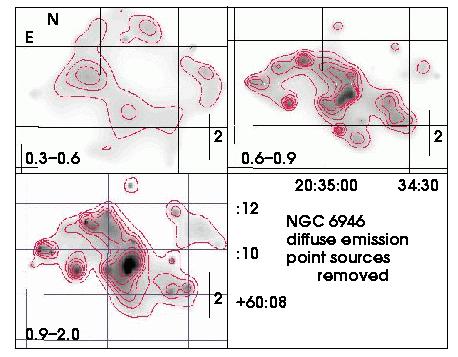~~~\label{mosaic}}
\end{center}
\end{figure*}

We then adaptively smoothed the data using the {\tt CIAO} code {\tt
csmooth}.  We adopted a gaussian smoothing kernel and selected the
smoothing parameter to be 3${\sigma}$ as calculated locally.  Finally,
we subtracted a constant from the result to remove the overall
background as determined by the `blank field' background.
Figure~\ref{mosaic} shows the results in the three bands.  The 0.3-0.6
keV band shows little emission specific to NGC 6946 because of the
high column in the galaxy's direction.  The 0.6-0.9 and 0.9-2 keV
bands appear similar: contours fall in similar regions of the galaxy
in each band.  In the 0.9-2 keV image, the apparent emission that
extends beyond the region defined by the 0.6-0.9 keV image is an
artifact of the adaptive smoothing, driven by the lower counts in the
0.9-2 keV band.  We henceforth use just the 0.6-0.9 keV band for any
comparisons.  The ``point sources'' apparent in this figure do {\it
not} correspond with any of the detected sources listed in
\cite{Holt02}, nor do they correspond with point sources lying below
the threshold.  An examination of the raw data at full resolution
shows a broad distribution of counts suggestive of a diffuse but
clumpy source.

\section{Multiwavelength Comparison}

In this section we compare the X-ray image of the diffuse emission
with images obtained in other bandpasses.  

The apparent ``point sources'' discussed at the end of the previous
section do not correspond to any of the optically-identified supernova
remnants \citep{MF97}.  The regions may correspond with prominent H~II
regions identified in the submillimeter band by \cite{Alton02}.  The
extracted counts from these regions yield luminosities of
$\sim$1.5$\times$10$^{37}$ erg s$^{-1}$ using a simple thermal
bremsstrahlung model with temperature of 0.3 keV (\S4.4).

\begin{figure*}
\begin{center}
\caption{X-ray contours for 0.6-0.9 keV overlaid on an H${\alpha}$
image; the H${\alpha}$ image is courtesy of Robert Fesen \citep{MF97}.  
The X-ray contours are described in Figure~\ref{mosaic}.~~~FIGURE IN
SEPARATE JPG FILE:  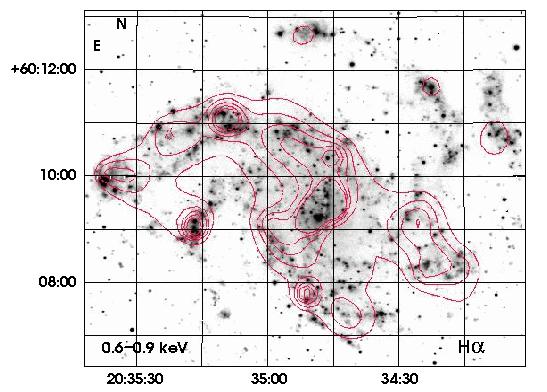~~~ \label{figXHalf}}
\end{center}
\end{figure*}

Figure~\ref{figXHalf} shows the contours of the 0.6-0.9 keV emission
overlaid on an H${\alpha}$ image.  Note that in general the X-rays
follow the H${\alpha}$ emission.  This behavior is most easily visible
on the prominent northern arm or at the sharp south-southeast edge.
The weak arm to the northwest is not detected; nor is the weaker
portion of the inner eastern arm.  The matching behavior means that
the X-ray emission arises from a hot diffuse ISM component or from the
summed emission of point sources confined to the arms.  We will return
to this question in the discussion section.

\begin{figure*}
\begin{center}
\caption{X-ray contours for 0.6-0.9 keV overlaid on 6 cm Polarized
Intensity map from \cite{BH96}.  Note the contours lie between the
`polarized spiral arms' as described by \cite{BH96}. The vertical
bar is 2$'$ in length.  ~~~ FIGURE IN SEPARATE JPG FILE: 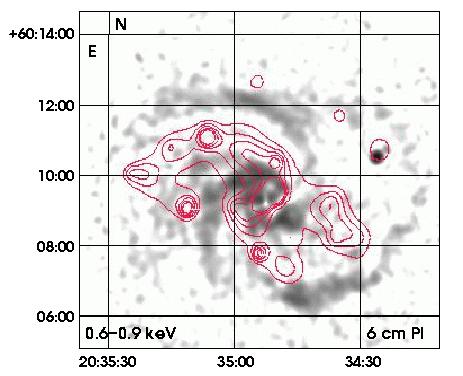 ~~~
\label{figXPI}}
\end{center}
\end{figure*}

Figure~\ref{figXPI} shows the 0.6-0.9 keV X-ray contours over the 6-cm
polarized intensity map from \citep{BH96} and discussed in
\cite{Frick}.  \cite{BH96} showed that the regular magnetic field is
strongest in the {\it interarm} region.  In the previous figure, we
established that the X-ray emission follows the northern arm. As in
\cite{BH96}, it is clear the polarized emission lies external to that
arm, so the X-ray image is consistent with their result.

\begin{figure*}
\begin{center}
\caption{ SCUBA 850 $\mu$ image with contours of X-ray diffuse
emission (0.6-0.9 keV) overlaid.  The 850 $\mu$m data are from
\cite{Alton02}.  The gray levels have been set to match approximately
the contours defined by Alton et al. in which the lowest visible gray
level corresponds to their 3${\sigma}$ isophote.  The vertical bar is
2$'$ in length. ~~~ FIGURE IN SEPARATE JPG FILE:  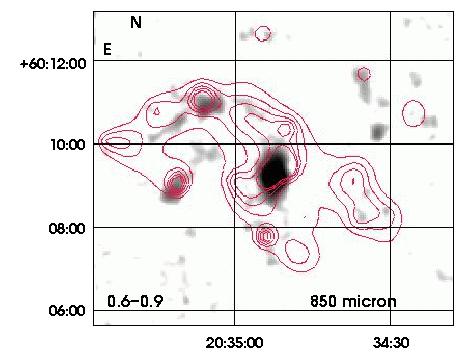~~~\label{figX850}}
\end{center}
\end{figure*}

Figure~\ref{figX850} shows the 0.6-0.9 keV X-ray contours overlaid on
the 850 ${\mu}$m map of \cite{Alton02}.  At 850 ${\mu}$m, the nucleus as
well as 3 knots of emission are prominent; the knots lie to the
northeast, east, and south-southeast of the nucleus.  Overall, the
regions of strongest X-ray emission correspond with the regions
brightest at 850 ${\mu}$m.  These knots correspond to prominent H II
regions.  The 850 ${\mu}$m emission largely arises from dust heated by
starlight and thermally radiating (e.g., \citealt{Is99}).

\begin{figure*}
\begin{center}
\caption{3 cm thermal radio emission with contours of X-ray diffuse
emission (0.6-0.9 keV) overlaid.  The 3 cm data are from
\cite{Frick}.  ~~~FIGURE IN SEPARATE JPG FILE:  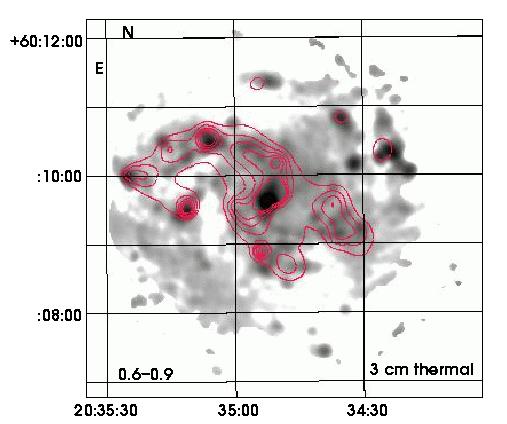~~~\label{fig3cm}}
\end{center}
\end{figure*}

Figure~\ref{fig3cm} displays the 0.6-0.9 keV X-ray contours on the 3
cm thermal radio emission (available in \citealt{Frick}).  In general,
knots of X-ray emission align with the radio knots, particularly on
the NE arm.  A longer X-ray exposure is required to ascertain whether
the outer regions in the 3 cm image have a corresponding X-ray
counterpart.

\section{Spectral Fit: Introduction}

For all of the analysis, we fit the spectra of the diffuse emission
$+$ back\-ground and the background simultaneously.  This introduction
describes the extraction of the spectra and a fit to the background.

Two approaches are generally used to extract a spectrum spread across
a large expanse of the ACIS chips: (i) extract the data as small
arrays, build a response matrix specific to each array, and sum
the arrays; or (ii) correct the entire chip for the charge transfer
inefficiencies (CTI), extract a single source$+$background spectrum,
and construct one response matrix (the Penn State approach, using the
CTI corrector as described in \citealt{Town02}).  We adopted the
second method.  A correction for the time-dependent contamination was
included by reducing the effective area at each affected
energy\footnote{Information about the QE degradation is available at
http://asc.harvard.edu/cal/Acis/Cal\_prods/qeDeg/index.html.}.

Following the removal of the point sources, the spectrum of the
diffuse emission$+$background was extracted including all pixels
within the S3 chip and with exposure $>$0.9 of the on-axis time (this
eliminates edge effects).  A total of 28248 counts were extracted
representing the spectrum of the source plus background.  A background
spectrum was extracted from the reprojected blank field observation
using the same (x, y) image pixels as for NGC 6946 itself.  The
background spectrum contained 331600 counts and 22334 counts after
scaling the exposure time to that of NGC 6946.  The source spectrum
was adaptively binned so that the resulting spectrum contained at
least 20 counts per channel.

\begin{figure*}
\begin{center}
\scalebox{0.35}{\rotatebox{-90}{\includegraphics{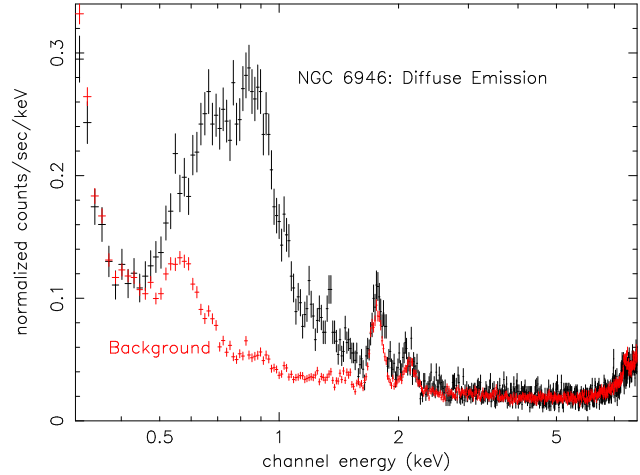}}}
\caption{Plot of the diffuse + background (upper curve) and background
(lower) spectra.  The background spectrum was extracted from a blank
sky observation after reprojecting those events using the aspect
solution from the NGC 6946 observation and scaling by the ratio of the 
exposure times between the background and the NGC 6946 observations.  
The diffuse $+$ background spectrum contains 28248 counts; the background
spectrum contains 22334 counts.  The difference is the net spectrum of 
the diffuse emission and contains 5914 counts.   \label{figovrly}}
\end{center}
\end{figure*}

Figure~\ref{figovrly} shows the two extracted spectra as an overlay.
From this figure, it is clear that the galaxy's diffuse emission fades
into the background near 2 keV and is blocked below $\sim$0.5 keV by
the column toward NGC 6946.  This figure reinforces one's visual
impression as well as quantitative expectation: most of the visually
apparent emission appears in the 1-2 keV band where {\it Chandra}'s
effective area takes on the largest values.  The visible `emission
lines' arise from particle background stimulation of material in the
detectors; the large line at 1.8 keV is Si K${\alpha}$
\citep{PropGuide}\footnote{The online version shows the background
spectrum in red, the source+background spectrum in black; from the
figure, it is clear the source spectrum contains excess 1.8 keV
emission above the instrumental feature}.  The difference between the
two spectra of Figure~\ref{figovrly} represents the net diffuse
spectrum and contains 5914 counts.

\begin{figure*}
\begin{center}
\caption{Fitted background spectrum using a combination of power law
and gaussian model components.  This figure may be compared to the
corresponding figure in \cite{PropGuide}.  The residuals are flat
to within $\sim$15\%. ~~~FIGURE IN SEPARATE JPG FILE: 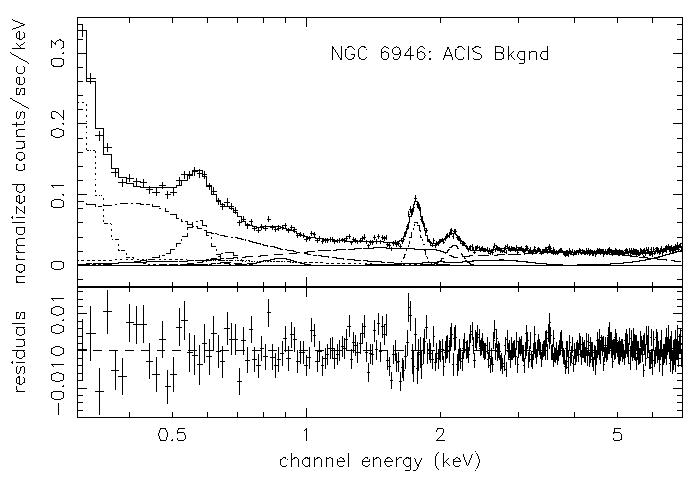~~~
 \label{spec_bkgrd}}
\end{center}
\end{figure*}

Figure~\ref{spec_bkgrd} shows a fit to the background using a number
of power law and gaussian components.  The residuals are flat to
within $\sim$15\% across the bandpass.  We use this background model
in all subsequent spectral fitting, freezing the parameters at their
best-fit values.  We justify this approach by noting the background
spectrum may be fitted to very high precision, it having been
accumulated from numerous long exposures.

\subsection{Fit to All Diffuse Emission} \label{alldiff}

Simple continuum models did not provide a good fit and in all cases
left clear residuals.  For example, the use of a single Mekal model
produced the lowest ${\chi}^2/{\nu}$ of 1.21 of all the {\it
single}-component models, but left clear residuals in the 0.9 keV
band.  The residuals essentially force consideration of additional
degrees of freedom.

\begin{figure*}
\begin{center}
\caption{(top) Spectral fit to the diffuse + background spectrum using
a variable dual Mekal plasma model. The instrumental background has
been fitted using the model of Figure~\ref{spec_bkgrd}.  The
individual model components are also shown.  (bottom) Spectral fit to
the diffuse data using a thermal bremsstrahlung and a series of
gaussians.  The fitted bremsstrahlung temperature is identical, within
the errors, to the temperature of the top figure.  ~~~ FIGURES IN
SEPARATE JPG FILES: 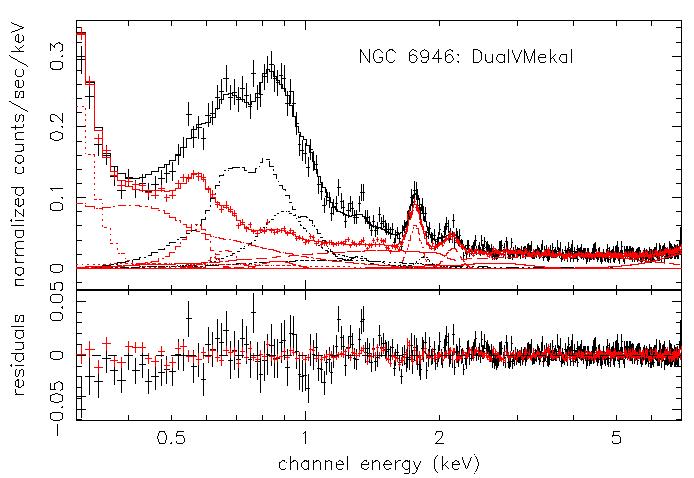, 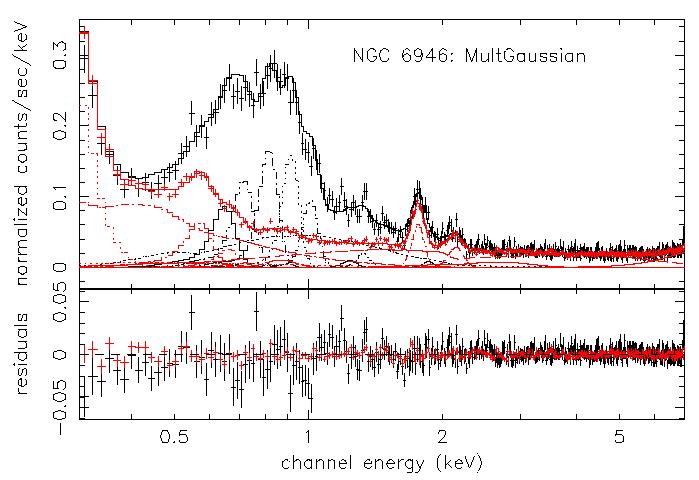~~~\label{figspec}}
\end{center}
\end{figure*}

Four models provided essentially comparable goodness-of-fit values.
At the interpretation stage, we will use the results from one or more
models because any one of them could be deemed "correct."  First, we
adopted a double `Mekal' spectral model representing thermal emission
from a diffuse gas as calculated by \cite{Mewe85,Mewe86} and
\cite{Kaa92} with Fe L calculations supplied by \cite{Lied95}.  Second
(and third), we used single and dual variable abundance Mekal models,
allowing the abundances of species with lines in the 0.5-2 keV band to
vary.  Last, we fit the spectrum with a simple thermal bremsstrahlung
to define the continuum and included as many gaussian lines as
required by the data.  For the multi-gaussian approach, the choice of
continuum model is largely irrelevant, as any simple model will
produce a fit.  A power law, for example, works as well as a
bremsstrahlung model (the resulting index is a steep $\sim$7$\pm$0.55,
90\% error); we adopt the bremsstrahlung model to provide a check on
the fitted temperatures from the more complex models.  For the
gaussian lines, we set their widths to zero, because the lines are
unresolved with ACIS but broadened by the instrumental response, and
their centers to the positions of expected lines
(\S\ref{secLineEmis}).  The multi-gaussian approach provides
information on the ionization states contributing to the total
emission.  Figure~\ref{figspec} (top) shows the dual variable Mekal
fit while the bottom figure shows the multi-gauss fit.

We also included a power law component to account for any excess
emission between the source plus background and background spectra.
The residual flux from this power law `background' component is
$\lesssim$6\% of the total diffuse flux, verifying the quality of the
background definition and its separation from the diffuse emission of
NGC 6946.  Table~\ref{specfit} lists the parameters for the fitted
models.

\subsection{Continuum Emission} \label{secContEmis}

\begin{figure*}
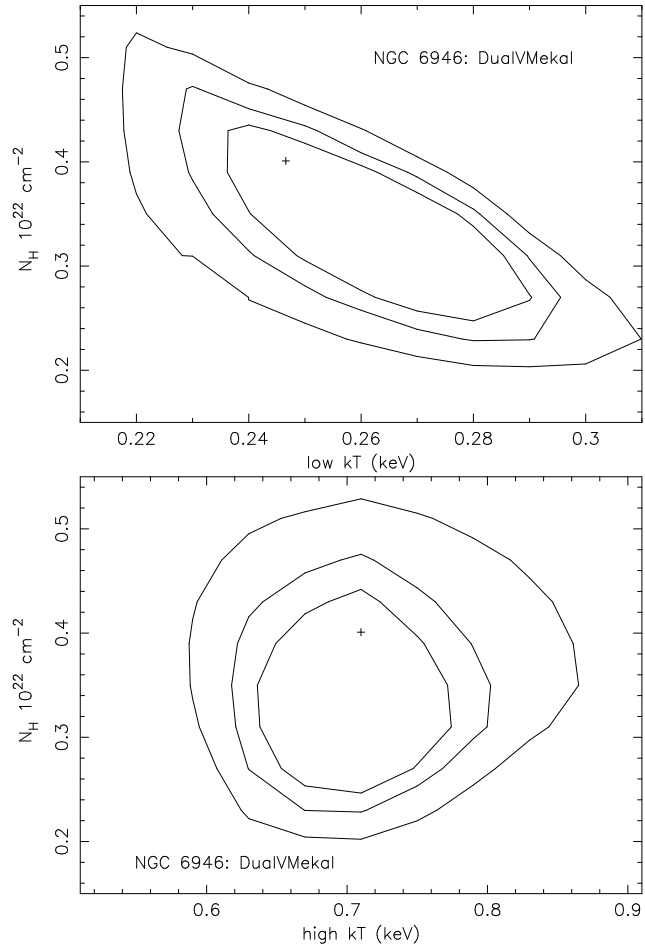

\begin{center}
\scalebox{0.35}{\rotatebox{-90}{\includegraphics{f10a.eps}}}
\scalebox{0.35}{\rotatebox{-90}{\includegraphics{f10b.eps}}}
\caption{The contour plot for the variable Mekal models of the errors
on the fitted parameters of the two thermal temperatures and the column.
The contours for this and the next figure are 1 ${\sigma}$, 90\%, and
99\% for 2 parameters of interest (${\Delta}{\chi}^2$ = 2.30, 4.61,
and 9.21).  \label{cont_vmekal}}
\end{center}
\end{figure*}

The continuum temperatures of the single continuum models or the {\it
low} temperature component of the dual models are identical within the
errors with kT $\sim$0.25-0.32 keV.  Figure~\ref{cont_vmekal} (top)
shows the variable Mekal model contours for the low-temperature
component.  The fitted column density is also identical within the
errors for all models with a range of 3.6-4.1$\times$10$^{21}$
cm$^{-2}$.  These columns lie about a factor of 2 above the Dickey \&
Lockman column ($\sim$2$\times$10$^{21}$ cm$^{-2}$, \citealt{DL90}) but within
the errors of the \cite{SFD} value of $\sim$3-5$\times$10$^{21}$ cm$^{-2}$.

The integrated flux in the 0.5-2 keV band is $\sim$3.7$\times$10$^{-13}$
ergs s$^{-1}$ cm$^{-2}$ (absorbed) or $\sim$2.5$\times$10$^{-12}$
(unabsorbed).  The integrated luminosity values are
$\sim$1.5$\times$10$^{39}$ and $\sim$1$\times$10$^{40}$ erg s$^{-1}$,
respectively, or $\sim$2.9$\times$10$^{37}$ and
$\sim$1.9$\times$10$^{38}$ erg s$^{-1}$ arcmin$^{-2}$, respectively.

These values differ somewhat from previous measurements.  The {\it
ROSAT} measure of the column was low by about a factor of 2
\citep{Sch94}.  The fitted temperature is also lower than the {\it
ROSAT} PSPC value of 0.55$^{+0.10}_{-0.25}$ keV; even though the error
bars just overlap, the higher temperature behavior of the {\it ROSAT}
fit is not unexpected.  The fitted temperature, for example, from the
PSPC data was almost certainly artificially high because it included
photons from sources with harder spectra, such as X-ray binaries, that
were incompletely removed.  In addition, \S\ref{alldiff} shows that
two spectral components are required for a good fit with one at kT
$\sim$0.7 while the PSPC data were consistent with a single
temperature because of the detector's lower spectral resolution.  The
integrated flux is higher than the {\it ROSAT} value by $\sim$25\%, as
is expected given the higher value for the column density.  As an
aside, the very broad PSF of the {\it ASCA} mirrors prohibited an
analysis of the diffuse emission from the NGC 6946 observation
\citep{SBF00}.

\subsection{Line Emission and Abundances} \label{secLineEmis}

\begin{figure*}
\begin{center}
\scalebox{0.35}{\rotatebox{-90}{\includegraphics{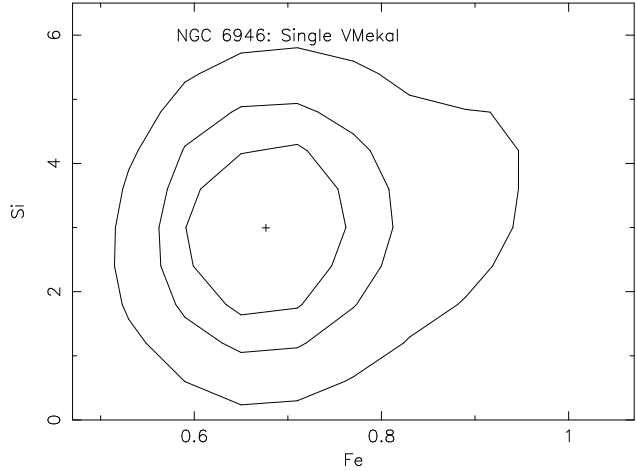}}}
\caption{Contours for the abundances of Fe and Si from the variable
Mekal spectral fit.  The abundance of Si differs from solar at the
90\% level but not at 99\%.  The Fe abundance differs at 99\%.
\label{mek_cont}}
\end{center}
\end{figure*}

The abundances of the single variable Mekal model were first varied
individually to test the size of the resulting error.  An abundance,
for which the 1 ${\sigma}$ error was consistent with 1.0, was reset to
and fixed at 1.0.  Two abundances (Si, Fe) were found to differ
significantly from 1 (Figure~\ref{mek_cont}).  For Fe, the abundance
is 0.67$\pm$0.13 solar at the 99\% level.  For Si, the abundance is
3.0 with the 90\% contour just above 1.0 at 1.05; at 99\%, the Si
abundance is consistent with solar.  No flares occurred during the
observation, so the line is unlikely instrumental in origin.  We
return to this point momentarily.

The gaussian centers were fixed at the energies of potentially
prominent lines, corresponding to O~VII 0.57 keV, O~VIII 0.653 keV,
Fe~I (L) 0.705 keV, Fe~XVII 0.826 keV, Ne~IX 0.922 keV, Fe~XX 0.996
keV, Ne~X 1.022 keV, and Mg~XI 1.34 keV.  Lines at energies lower than
$\sim$0.5 keV were deemed irrelevant because of the high column toward
NGC 6946.  Several lines were significantly detected at $>$99\%
(Table~\ref{specfit}).  At the energies of these lines, the ACIS spectral
resolution is $\sim$110-120 eV \citep{PropGuide}.

While we fixed the line centers at specific energies to test for the
presence or absence of known lines, we relaxed that requirement for
the Si line to test whether the feature in the source$+$background
spectrum was consistent with the background feature.  The background
line\footnote{Recall that the background spectrum may be arbitrarily
precise because it is an accumulation of an arbitrary number of blank
sky events.} is best fit with a gaussian at 1.775$\pm$0.002 keV (blend
of Si K${\alpha}$) while the diffuse Si line has a best-fit energy of
1.860$\pm$0.027 keV (= Si XIII).

Assume for the moment the spectral fits are accurate measures of the
line emission.  The presence of O~VII and O~VIII then provides a
measure of the overall ionization in the diffuse emission.  O~VIII
0.653 keV is the Ly$\alpha$ transition; the blended O~VII 0.57 line(s)
represent the 1s2p to 1s$^2$ transition.  The log of the ratio is
-0.2$\pm$0.7 and has a large error because of the decreasing continuum
in the 0.5-0.6 keV band and because of the energy resolution of the
CCD.  Nevertheless, the ratio translates to limits of 9.6 $\lesssim$
log n$_e$t $\lesssim$12.0, where n$_e$ represents the electron density
and t the cooling time (see, for example, \citealt{Ved86}).  If the
low `vmekal' continuum temperature approximately represents the
electron temperature, then it plus the line ratio restrict the O~VII
resonance/forbidden ratio to $\lesssim$0.6-0.8, a prediction for
future observations using instruments with higher energy resolution.
The values are just barely consistent with ionization equilibrium and
suggest non-equilibrium conditions exist in the diffuse medium of NGC
6946 particularly given the lack of a good fit by the various
equilibrium ionization models.  For the just-described electron
temperature and line ratio values, equilibrium requires n$_e$t
$\gtrsim$12.5-13.0.  Lower values for the electron temperature plus
lower values for the interpreted line ratio push the estimated
resonance/forbidden ratio toward higher values and increasing
equilibrium conditions; for a given line ratio, higher electron
temperatures indicate increasing non-equilibrium conditions. 

A recent paper by \cite{Page03} presents the {\it XMM-Newton} RGS
spectrum of gas in M81.  We note that the spectrum shows the same
resolved oxygen emission lines with a ratio of $\sim$0.65, indicating
non-equilibrium conditions exist in the ISM of M81.  While we
acknowledge that the diffuse emission in NGC 6946 requires study with
an instrument possessing higher spectral resolution at a spatial
resolution comparable to {\it Chandra}'s to confirm our analysis, the
similarity between the line ratio in the RGS spectrum and our upper
limit argues for non-equilibrium conditions in the ISM of both
galaxies.

A similar analysis using Ne~IX and Ne~X can not be carried out because
the neon emission is blended with Fe XIX and Fe XX.  The M81 spectrum
of Page et al. in fact illustrates that the blend occurs even at the
resolution of the RGS.

A deeper X-ray exposure will be necessary to provide sufficient events
to tighten the line ratio and continuum temperature values to
strengthen the non-equilibrium suggestion.  A future mission with high
spatial and spectral resolution is necessary to demonstrate
conclusively whether the diffuse emission exists in a non-equilibrium
condition by measuring the resonance/forbidden ratio of the O VII
lines.  The expected large point spread function of {\it Astro-E2}
will blur too many point sources into the diffuse emission to provide
such a good measure.

\subsection{Diffuse Emission without the Knots and the Spectrum of the Knots}

\begin{figure*}
\begin{center}
\scalebox{0.35}{\rotatebox{-90}{\includegraphics{f12b.eps}}}
\caption{(upper) Spectral fit, using a thermal bremsstrahlung and a
series of gaussians, to the diffuse data minus the knots of emission.
The fitted bremsstrahlung temperature is identical to the
bremsstrahlung temperature of figure~\ref{figspec} within the errors.
(lower) Spectral fit to the emission of the knots using a
bremsstrahlung and three gaussians.  Because of the small spatial
extent of the knots, the background is nearly negligible and has not
been shown in the figure; it was included in the spectral fitting.
~~~ FIGURE f12a IN SEPARATE JPG FILE: 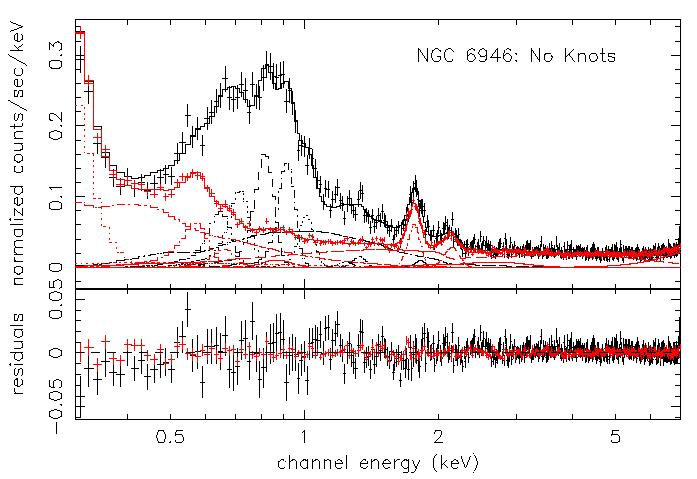~~~\label{fignoknot}}
\end{center}
\end{figure*}

As an experiment, we extracted a spectrum of the diffuse emission in
the knots visible in the smoothed images as well as a spectrum of the
diffuse emission minus the knot emission.  Figure~\ref{fignoknot}
shows the results.  The spectrum of the `no knot' diffuse emission is
essentially identical to the complete spectrum discussed previously.
This is not particularly surprising given the relatively bright
emission distributed over the face of the galaxy and the relatively
small number of events in the knots.

The lower portion of the figure shows the knot emission itself.  We
fit the spectrum to obtain the column and continuum temperature with a
single variable Mekal model, then fit for line abundances as well as
using gaussians to represent emission lines (as shown in the figure).
Table~\ref{specfit2} lists the model parameters.  The background forms
an inconsequential portion of the flux given the small spatial
coverage of the knots.  The abundances show enhanced Si but depleted
O, Ne, and Fe.  Gaussian line centers correspond to O, Si, and Fe; the
fourth line at $\sim$0.95 keV is most likely a blend of Ne IX and Fe
XX.  The integrated flux of the knot emission is
$\sim$3$\times$10$^{-13}$ erg s$^{-1}$ cm$^{-2}$ or about 5--10\% of
the original diffuse emission.  

\section{Discussion}

The spectra of the point sources with sufficient counts to yield good
spectral fits, ${\chi}^2 / {\nu}$ $\lesssim$1.3 for ${\nu} \gtrsim$50,
are uniformly harder than the diffuse spectrum presented here
\citep{Holt02}.  The lowest temperature of the best-fit bremsstrahlung
models for the point sources is $\sim$1.9 keV.  Similarly, the
best-fit power law indices are typically $\sim$2-3; the steepest value
is 4.9$\pm$0.4 versus the power law index fit to the diffuse emission
of 7$\pm$0.55 (\S\ref{alldiff}).  These results all support the
presence of a distinct spectral component distributed across the
galaxy.

\cite{Sch94} argued that the detection of the diffuse emission
represented the hot ISM component and not the summed emission of point
sources based upon a plausibility argument from the known luminosities
of typical constituents.  The argument is strengthened here because of
the improved separation of point sources and diffuse emission as well
as the deeper probe of the luminosity distribution of the {\it
Chandra} observation.  \cite{Holt02} detected 72 point sources to
$\sim$7$\times$10$^{36}$ erg s$^{-1}$ and judged it complete to
$\sim$10$^{37}$ erg s$^{-1}$ which is a factor of $\sim$10 deeper than
the PSPC data and a factor of $\sim$7 deeper than the HRI image
(\citealt{Sch94,SBF00}).

Our argument can be strengthened still further by extrapolating the
luminosity distribution to an arbitrary, low value of L$_{\rm X}$.  We
adopt the luminosity distribution of Holt et al. but eliminate from
that distribution any sources not detected in the 0.5-2.0 keV band.  A
total of 7 sources are removed.  A fit to the resulting distribution
changes the distribution's slope marginally from 0.64$\pm$0.02 to
0.62$\pm$0.01.  If we extrapolate this distribution to L$_{\rm X}$
$\sim$10$^{28}$ erg s$^{-1}$, we obtain a summed L$_{\rm X}$ of
$\sim$3.3$\times$10$^{38}$ erg s$^{-1}$; pushing the distribution fit
parameters to their positive error limits yields a summed L$_{\rm X}$
of $\sim$3.6$\times$10$^{38}$ erg s$^{-1}$.  When subtracted from the
fitted diffuse luminosity, we obtain the lower bound on the diffuse
luminosity, $\sim$1.1$\times$10$^{39}$ erg s$^{-1}$.  Of the events
described in this paper as `diffuse' then, up to $\sim$25\% could
originate in unresolved point sources below the detection threshold,
leaving $\sim$75\% for the hot component of the ISM of NGC 6946.

The above discussion assumes the lack of a yet-steeper contribution
from the unresolved point sources.  We can not assess the impact of
such a component but note in defense of our argument that the
luminosity distribution of the Milky Way does not contain evidence for
the existence of such a component \citep{Grimm02}.

The unresolved point source contribution may very well be less than
our 25\% estimate.  \cite{Grimm02} estimate the number of sources
above 10$^{34}$ erg s$^{-1}$ in the Milky Way as $\sim$700 with a
possible factor of 2 uncertainty.  For NGC 6946, our extrapolation
leads to $\sim$3800 sources above 10$^{34}$ erg s$^{-1}$, a factor of
5 greater than the Milky Way.  If the extrapolation for NGC 6946 lies
closer to the Milky Way observations, the unresolved point
contribution likewise drops, perhaps by as much as a factor of 2.  Of
course, this argument could equally well be pushed in the opposite
direction by a similar factor because NGC 6946 is not a clone of the
Milky Way.  For example, NGC 6946 is extremely CO-rich, radio-bright,
and undergoing very active star formation \citep{Cas90,Walsh02}.  If
we doubled the unresolved point source contribution, our estimate for
the hot ISM contribution decreases to $\sim$50\% of the X-rays
presented here as the `diffuse spectrum' for a luminosity of
$\sim$7$\times$10$^{38}$ erg s$^{-1}$.

Noting the caveats above but taking the original measurement at face
value, our results represent a solid detection of the hot diffuse
component from NGC 6946 free of any concerns over the resolved
sources' point spread function wings that plagued results from {\it
ROSAT}, for example \citep{Sch94}.  Summing the unresolved point
component, the diffuse component (with a luminosity in the range of
$\sim$50-75\% of the diffuse L$_{\rm X}$ discussed in this paper), and
the resolved sources \citep{Holt02} yields a total L$_{\rm X}$ of
$\sim$1.3$\times$10$^{40}$ erg s$^{-1}$ in the 0.5-2 keV band.  The
hot diffuse component then represents at least $\sim$5-8\% of the
total X-ray emission of NGC 6946 and potentially as much as
$\sim$10\%.

If we approximate NGC 6946, based on the 0.6-0.9 keV image, as a thin
rectangle of length $\sim$9$'$.5, width $\sim$4$'$.5, and thickness
h$_{\rm kpc}\times$1 kpc, the resulting volume V is
$\sim$3.8$\times$10$^{66}$ h$_{\rm kpc}$ cm$^3$.  We adopt 1 kpc for
the unit of thickness for convenience.  The X-ray-emitting gas may be
distributed in a layer of different thickness.  For example,
\cite{EB93} found a consistent model for thermal radio emission and
Faraday rotation with a thin disk, but argued for a thicker disk.
They did so based upon the assumption that Faraday rotation and
depolarization occurred in the identical volume, assumptions not
necessarily valid\footnote{We thank the referee for drawing our
attention to this point.} because edge-on galaxies show both thin and
thick disks in the thermal gas and synchrotron emission.

The continuum model normalization parameter is proportional to the
emission measure $\mathcal{E}$ = ${\int} n_{\rm e}^2 {\rm dV}$.  From
the measured spectral temperature (the low temperature component) and
the emission measure, we may estimate additional parameters describing
the ISM following \cite{Wang95} or \cite{Summ03} to be consistent with
their definitions.  For a filling factor f and proton mass m$_{\rm
p}$, the estimated gas density $n_{\rm e}^2$ =
$\mathcal{E}$~/~(fVh$_{\rm kpc}$) or n$_{\rm e} {\sim}$0.012 (fh$_{\rm
kpc}$)$^{-\frac{1}{2}}$ cm$^{-3}$.  The mass of the gas in the diffuse
component is M $\sim$ fn$_{\rm e}$ V m$_{\rm p}$
$\sim$3.8$\times$10$^7$ (${{\rm h}_{\rm kpc}}{\rm
f})^{\frac{1}{2}}$M$_{\odot}$.  The gas pressure p = 2 n$_{\rm e}$ kT
= 9.1$\times$10$^{-12}$ (fh$_{\rm kpc}$)$^{-\frac{1}{2}}$ dyn
cm$^{-2}$, the thermal energy E $\sim$3 n$_{\rm e}$ kT V $\sim$
5.2$\times$10$^{55}$ ($\frac{{\rm h}_{\rm kpc}}{\rm
f}$)$^{\frac{1}{2}}$ erg, and the cooling time is t$_{\rm cool}$
$\sim$3kT/${\Lambda}n_{\rm e}$ $\sim$1.9$\times$10$^7$ (fh$_{\rm
kpc}$)$^{\frac{1}{2}}$ yr, with ${\Lambda}$ = 1.6$\times$10$^{-23}$
erg cm$^3$ s$^{-1}$ adopted for the gas emissivity \citep{SuthDop93}.
The mass of X-ray emitting gas represents $\sim{0.3 \frac{h_{\rm
kpc}}{\rm f}}$\%, for a filling factor f, of the total molecular gas
mass (H$_2$ $\sim$10$^{10}$ M$_{\odot}$, \citealt{YS82}).  The cooling
rate $\dot M_{\rm cool}$ $\sim{{M_{\rm X}} \over {t_{\rm cool}}}$
${\approx}{L_{\rm X} \over {T}}$ \citep{NSF84} and for the measured
values yields $\sim$2 M$_{\odot}$ yr$^{-1}$.  If we assume the
bremsstrahlung or Mekal temperatures represent the actual temperatures
of hot gas, they correspond to thermal velocities of 180-300 km
s$^{-1}$.  The number density is about the same, the thermal energy is
$\sim$5 times lower, the pressure $\sim$20\% lower, the X-ray-emitting
gas mass $\sim$6 times higher, and the cooling time about 2 times
shorter, than the corresponding values for NGC 4449 based upon its
{\it Chandra} observation \citep{Summ03}.  How these values correlate
with other galaxian parameters requires a considerably broader survey
of diffuse emission.  At this point, we note that NGC 6946 has a
higher total mass, more H I gas, and a higher 60 ${\mu}$m luminosity
\citep{Young89}, indicating more vigorous star formation, than NGC
4449 possesses.

The analysis of the RGS spectrum of M81 \citep{Page03} finds a dual
temperature plasma\footnote{Page et al. formally require a
contribution from a third component with kT$_3$ = 1.7$^{+2.1}_{-0.5}$
keV.  They argue that this component arises for the most part in the
bulge of M81.} describes most of the spectrum with kT$_1$ =
0.18$\pm$0.04 and kT$_2$=0.64$\pm$0.04 keV, very similar to our
results achieved at a lower spectral but higher spatial resolution.
They infer a cooling time of $\sim$4$\times$10$^7$ $\sqrt{f}$ yr,
similar to our inferred value.  \cite{Kun03} obtain a best-fit dual
temperature model of kT=0.20 and 0.75 keV for the diffuse component in
M101.  Finally, \cite{Ehle98} fit the diffuse component of M83 with a
dual temperature model and obtained kT$_1$ $\sim$0.2 and kT$_2$
$\sim$0.5 keV.  These results all support the existence of at least
a 2-temperature hot ISM in spiral galaxies.  

The X-ray knots correspond spatially with local maxima in the 850
${\micron}$ image.  The 850 ${\micron}$ maxima are knots of dust
heated by starlight and correspond with H~II regions or areas of star
formation activity \citep{Alton02}.  That the X-ray image shows local
maxima in the same location raises the possibility that the knots are
blowout regions (e.g. \citealt{Ott01}).  \cite{Walsh02} showed that
the total 6 cm radio image correlated with images from the mid-IR and
argued that the apparent single correlation was in fact two
correlations, between the clumpy mid-IR vs. 6 cm thermal radio and the
diffuse mid-IR vs. 6 cm nonthermal radio emission.  The diffuse radio
emission is dominated by the nonthermal component; the radio emission
from clumps was largely thermal in origin.  The correlation with the
X-ray then suggests the X-ray emission is thermal in origin, so line
emission should be present.  The knot emission shows at best weak
evidence for line emission within the statistical uncertainties.  A
higher signal-to-noise spectrum is certainly needed to improve or
refute the upper limits on the presence of X-ray line emission.  We
intend to pursue additional observations, particularly deeper
observations.

\acknowledgements

The authors thank the anonymous referee for suggestions that improved
the presentation of the paper.  This research was supported by
contract number NAS8-39073 to SAO.  The work made use of the {\tt
CIAO} software built under contract to SAO.  The authors thank
W. Walsh for providing the CO data, R. Beck for the 3, 6, and 6 cm PI
data, R. Fesen for the H${\alpha}$ image, and F. Combes for the 850
${\mu}$m data.

\newpage

\begin{table*}
\begin{center}
\caption{Fitted Parameters for the Spectrum of the Diffuse Emission\tablenotemark{a}}
\label{specfit}
\begin{tabular}{lllllll} \cr
Model\tablenotemark{b} & Parameter & N$_{\rm H}$\tablenotemark{c} & Norm  & Flux\tablenotemark{d}~or EqW & ${\chi}^2$/${\nu}$ & ${\nu}$\tablenotemark{e} \cr \hline
 Mekal (kT$_1$) +              & 0.255$^{+0.030}_{-0.025}$ & 0.36$^{+0.10}_{-0.08}$ & 9.4e-4 & 2.3$\pm$0.4e-12 & 1.22 & 351 \cr
 Mekal (kT$_2$) +              & 0.70$^{+0.11}_{-0.08}$ & $\cdots$ & 1.3e-4 & 1.8$\pm$0.6e-13 & $\cdots$ & $\cdots$ \cr
 ~~Power Law (Bkg) ($\Gamma$)    & 2.10$^{+0.48}_{-0.45}$ & $\cdots$ & 4.3e-5 & 1.0$\pm$0.3e-13 & $\cdots$ & $\cdots$ \cr
                           & \cr
Brems (kT) +               & 0.32$^{+0.30}_{-0.12}$ & 0.41$^{+0.10}_{-0.12}$ &  2.3e-2 & 2.3$\pm$0.5e-12 & 1.14 & 342 \cr
~~gaussian (norm, O VII) + & 0.569 & $\cdots$  & $\cdots$ & 403$^{+900}_{-220}$ & $\cdots$ & $\cdots$ \cr
~~gaussian (norm, O VIII) + & 0.653 & $\cdots$ & $\cdots$ & 266$^{+282}_{-203}$ & $\cdots$ & $\cdots$ \cr
~~gaussian (norm, Fe I L) + & 0.705 & $\cdots$ & $\cdots$ & $<$600 & $\cdots$ & $\cdots$ \cr
~~gaussian (norm, Fe XVII) + & 0.826 & $\cdots$ & $\cdots$ & 350$^{+90}_{-100}$ & $\cdots$ & $\cdots$ \cr
~~gaussian (norm, Ne IX) + & 0.922 & $\cdots$ & $\cdots$ & 289$^{+168}_{-120}$ & $\cdots$ & $\cdots$ \cr
~~gaussian (norm, Fe XX) + & 0.996 & $\cdots$ & $\cdots$ & $<$100  & $\cdots$ & $\cdots$ \cr
~~gaussian (norm, Ne X) + & 1.022 & $\cdots$  & $\cdots$ & 90$^{+80}_{-30}$ & $\cdots$ & $\cdots$ \cr
~~gaussian (norm, Mg XI) + & 1.340 & $\cdots$ & $\cdots$ & 60$^{+50}_{-20}$ & $\cdots$ & $\cdots$ \cr
~~gaussian (norm, Si XIII) + & 1.860 & $\cdots$ & $\cdots$ & 250$^{+160}_{-90}$ & $\cdots$ & $\cdots$ \cr
~~Power Law (Bkg) ($\Gamma$)      & 2.24$^{+0.40}_{-0.25}$ & $\cdots$ & 4.5e-5 & 9.5$\pm$0.4e-14 & $\cdots$ & $\cdots$ \cr  
 \cr
 VarMekal (kT$_1$) + & 0.245$^{+0.050}_{-0.015}$ & 0.40$^{+0.07}_{-0.17}$ & 1.3e-3 & 2.5$\pm$0.6e-12 & 1.19 & 351 \cr
 VarMekal (kT$_2$) + & 0.71$^{+0.10}_{-0.08}$ & $\cdots$  & 1.1e-4  & 2.8$\pm$0.8e-13 & $\cdots$ &$\cdots$ \cr
  ~~abundance: Si  + & 2.26$^{+1.44}_{-1.19}$ & $\cdots$ & $\cdots$ & $\cdots$ & $\cdots$ & $\cdots$ \cr
  ~~Power Law (Bkg) ($\Gamma$)   & 2.02$^{+0.53}_{-0.32}$ & $\cdots$ & 4.2e-5 & 9.4$\pm$1.9e-14 & $\cdots$ & $\cdots$ \cr
\cr
 SingVarMekal (kT) + & 0.276$^{+0.017}_{-0.030}$ & 0.39$^{+0.06}_{-0.05}$ & 1.5e-3 & 2.4$\pm$0.6e-12 & 1.23 & 350 \cr
  ~~abundance: O + & 0.83$^{+0.20}_{-0.13}$ & $\cdots$ &  $\cdots$ & $\cdots$ & $\cdots$ & $\cdots$ \cr
  ~~abundance: Si + & 3.00$^{+1.95}_{-1.90}$ & $\cdots$ &  $\cdots$ & $\cdots$ & $\cdots$ & $\cdots$ \cr
  ~~abundance: Fe + & 0.67$^{+0.14}_{-0.12}$ & $\cdots$ &  $\cdots$ & $\cdots$ & $\cdots$ & $\cdots$ \cr
  ~~Power Law (Bkg) ($\Gamma$)  & 2.02$^{+0.33}_{-0.25}$ & $\cdots$ & 5.5e-5 & 1.2$\pm$0.2e-13 & $\cdots$ & $\cdots$ \cr  \hline
\end{tabular}
\tablenotetext{a}{Errors are all 90\%.  Upper limits are all 99\%.  The 'Bkg' power law is used to account for any residual background emission incorrectly subtracted.}
\tablenotetext{b}{All gaussian line widths were fixed to 0.0 because the lines are unresolved but broadened by the instrumental resolution; line centers are fixed at the listed values.}
\tablenotetext{c}{Units = 10$^{22}$ cm$^{-2}$.}
\tablenotetext{d}{Unabsorbed flux in 0.5-2.0 keV band with exponent indicated by e-XX = 10$^{-XX}$;
units = erg s$^{-1}$ cm$^{-2}$; EqW = Equivalent Width, units = eV.}
\tablenotetext{e}{The degrees of freedom represent the total number in
the {\it source} spectrum; the background degrees of freedom, 558,
have been subtracted from the total number of degrees of freedom
counted by the software.}
\end{center}
\end{table*}

\newpage

\begin{table*}
\begin{center}
\caption{Fitted Parameters for the Knots/No Knots Spectra\tablenotemark{a}}
\label{specfit2}
\begin{tabular}{lllllll} \cr
Model\tablenotemark{b} & Parameter & N$_{\rm H}$\tablenotemark{c} & Norm  & Flux\tablenotemark{d}~or EqW & ${\chi}^2$/${\nu}$ & ${\nu}$\tablenotemark{e} \cr \hline
Knots Emission:\cr
 ~Brems +     & 20$^{+0.08}_{-0.02}$ & 0.52$^{+0.08}_{-0.10}$ & 4.1e-3 & 4.0$\pm$1.8e-13 & 0.75 & 310 \cr
 ~~gaussian (O) + 0.652 & $\cdots$ & 1.1e-4 & 180$\pm$110 & $\cdots$ & $\cdots$ \cr
 ~~gaussian (Fe) + 0.822 & $\cdots$ & 2.7e-5 & 170$\pm$75 & $\cdots$ & $\cdots$ \cr
 ~~gaussian (Ne+Fe) + 0.95 & $\cdots$ & 7.4e-6 & 90$\pm$45 & $\cdots$ & $\cdots$ \cr
 ~~gaussian (Si) + 1.80 & $\cdots$ & 7.5e-7 & $<$500 & $\cdots$ & $\cdots$ \cr
\cr
 ~VMekal +    & 0.24$^{+0.05}_{-0.02}$ & 0.40$^{+0.07}_{-0.17}$ &  2.2e-5 & 2.5$\pm$0.9e-13 & 0.76 & 315 \cr
  ~~abundance: O + & 0.37$^{+0.45}_{-0.25}$ & $\cdots$ &  $\cdots$ & $\cdots$ & $\cdots$ & $\cdots$ \cr
  ~~abundance: Ne + & 0.39$^{+0.34}_{-0.28}$ & $\cdots$ &  $\cdots$ & $\cdots$ & $\cdots$ & $\cdots$ \cr
  ~~abundance: Si + & 2.86$^{+2.95}_{-1.78}$ & $\cdots$ &  $\cdots$ & $\cdots$ & $\cdots$ & $\cdots$ \cr
  ~~abundance: Fe + & 0.32$^{+0.50}_{-0.18}$ & $\cdots$ &  $\cdots$ & $\cdots$ & $\cdots$ & $\cdots$ \cr 
 \cr
No Knots Emission: \cr
 ~Brems +   & 0.28$^{+0.06}_{-0.02}$ & 0.50$^{+0.04}_{-0.06}$ &  7.05e-3 & 3.6$\pm$0.94e-12 & 1.25 & 347 \cr 
 ~~gaussian (norm, O~VIII ) + & 0.653 & $\cdots$ & 3.95e-4 & 265$\pm$145 & $\cdots$ & $\cdots$ \cr
 ~~gaussian (norm, Fe~I~L ) + & 0.707 & $\cdots$ & 3.69e-4 & 245$\pm$85 & $\cdots$ & $\cdots$ \cr
 ~~gaussian (norm, Fe~XVII ) + & 0.826 & $\cdots$ & 2.02e-4 & 215$\pm$70 & $\cdots$ & $\cdots$ \cr
 ~~gaussian (norm, Si~XIII ) + & 1.860 & $\cdots$ & 5.33e-6 & 440$\pm$170 & $\cdots$ & $\cdots$ \cr
 ~~Power Law (Bkg) & 2.09$^{+0.36}_{-0.27}$ & $\cdots$ & 3.4e-4 & 2.4$\pm$1.1e-13 & $\cdots$ & $\cdots$ \cr 
\cr
 ~VMekal + & 0.28$^{+0.03}_{-0.02}$ & 0.49$^{+0.05}_{-0.06}$ &  2.57e-3 & 4.3$\pm$0.94e-12 & 1.30 & 351 \cr
 ~~ abundance: Ne + & 0.74$^{+0.31}_{-0.15}$ & $\cdots$ &  $\cdots$ & $\cdots$ & $\cdots$ & $\cdots$ \cr
  ~~abundance: Si + & 3.95$^{+1.95}_{-1.90}$ & $\cdots$ &  $\cdots$ & $\cdots$ & $\cdots$ & $\cdots$ \cr
  ~~abundance: Fe + & 0.53$^{+0.95}_{-0.90}$ & $\cdots$ &  $\cdots$ & $\cdots$ & $\cdots$ & $\cdots$ \cr 
 ~~Power Law (Bkg) & 2.02$^{+0.37}_{-0.25}$ & $\cdots$ & 3.2e-4 & 2.4$\pm$1.1e-13 & $\cdots$ & $\cdots$ \cr  \hline
\end{tabular}
\tablenotetext{a}{Errors are all 90\%.  Upper limits are all 99\%.  The 'Bkg' power law is used to account for any residual background emission incorrectly subtracted.}
\tablenotetext{b}{All gaussian line widths were fixed to 0.0 because the lines are unresolved but broadened by the instrumental resolution; line centers are fixed as listed.}
\tablenotetext{c}{Units = 10$^{22}$ cm$^{-2}$.}
\tablenotetext{d}{Unabsorbed flux in 0.5-2.0 keV band with exponent indicated by e-XX = 10$^{-XX}$;
units = erg s$^{-1}$ cm$^{-2}$; EqW = Equivalent Width, units = eV.}
\tablenotetext{e}{The degrees of freedom represent the total number in
the {\it source} spectrum; the background degrees of freedom, 558,
have been subtracted from the total number of degrees of freedom
counted by the software.}
\end{center}
\end{table*}

\appendix

\begin{table}
\begin{center}
\caption{Model Parameters for Instrumental Background}
\label{bkgrd_fit}
\begin{tabular}{llll}
  Component & Parameter\tablenotemark{a} & Norm & Gauss ${\sigma}$ (eV) \cr \hline
 ~Power law & -0.223 & 3.70e-5 & $\cdots$ \cr
 ~Power law &  3.871 & 3.47e-5 & $\cdots$ \cr
 ~Power law &  2.789 & 6.77e-6 & $\cdots$ \cr
 ~Gaussian  &  0.670 & 1.03e-5 & 4.84e-4 \cr
 ~Gaussian  & 10.51  & 9.12e-3 & 1.642   \cr
 ~Gaussian  &  2.149 & 2.21e-5 & 5.64e-2 \cr
 ~Gaussian  &  0.577 & 4.09e-5 & 3.65e-2 \cr
 ~Gaussian  &  0.232 & 5.08e-3 & 4.77e-2 \cr
 ~Gaussian  &  1.775 & 1.83e-5 & 2.61e-2 \cr
 ~Gaussian  &  2.629 & 2.75e-5 & 0.468    \cr
 ~Gaussian  &  0.882 & 4.06e-6 & 5.06e-2 \cr  \hline
\end{tabular}
\tablenotetext{a}{for power law, parameter = ${\Gamma}$; for gaussian,
parameter = line energy}
\end{center}
\end{table}

\begin{thebibliography}

\bibitem[Alton et al.(2002)]{Alton02} Alton, P. B., Bianchi, S.,
Richer, J., Pierce-Price, D., and Combes, F. 2002, A\&A, 388, 446

\bibitem[Beck \& Hoernes(1996)]{BH96} Beck, R. \& Hoernes, P. 1996,
Nature, 379, 47

\bibitem[Casioli et al.(1990)]{Cas90} Casioli, F., Clausset, F.,
Combes, F., Viallefond, F., \& Boulanger, F. 1990, A\&A, 233, 357

\bibitem[CXC Proposers' Guide(2002)]{PropGuide} CXC Proposers'
Observatory Guide (2002), rev. 5 (Cambridge, MA: Chandra X-ray Center)

\bibitem[Della Ceca, R., Griffiths, R. E., \& Heckman,
T. M.(1997)]{Della97} Della Ceca, R., Griffiths, R. E. \& Heckman,
T. M. 1997, ApJ, 485, 581

\bibitem[Dickey \& Lockman(1990)]{DL90} Dickey, J. M. \& Lockman,
F. J. 1990, ARAA, 28, 215

\bibitem[Ehle \& Beck(1993)]{EB93} Ehle, M. \& Beck, R. 1993, A\&A,
273, 45

\bibitem[Ehle et al.(1998)]{Ehle98} Ehle, M., Pietsch, W., Beck, R.,
\& Klein, U. 1998, A\&A, 329, 39

\bibitem[Fabbiano(1989)]{F89} Fabbiano, G. 1989, ARAA, 27, 87

\bibitem[Fabbiano \& Trinchieri(1987)]{FT87} Fabbiano, G. \&
Trinchieri, G. 1987, ApJ, 315, 46

\bibitem[Freeman et al.(2002)]{Free02} Freeman, P. E., Kashyap, V.,
Rosner, R., \& Lamb, D. Q. 2002, ApJS, 138, 185

\bibitem[Frick et al.(2001)]{Frick} Frick, P., Beck, R., Berkhuijsen, E. M.,
\& Patrickeyev, I. 2001, MNRAS, 327, 1145

\bibitem[Grimm et al.(2002)]{Grimm02} Grimm, H.-J., Gilfanov, M., \&
Sunyaev, R. 2002, A\&A, 391, 923 

\bibitem[Hartline(1979)]{Har79} Hartline, B. K. 1979, Science, 205, 31

\bibitem[Holt et al.(2003)]{Holt02} Holt, S. S., Schlegel, E. M.,
Hwang, U., \& Petre, R.. 2003, ApJ, 588, 792

\bibitem[Israel et al.(1999)]{Is99} Israel, F. P., van der Werf, P. P.,
Tilanus, R. P. J. 1999, A\&A, 344, L83

\bibitem[Kaastra(1992)]{Kaa92} Kaastra, J.S. 1992, An X-Ray Spectral
Code for Optically Thin Plasmas (Internal SRON-Leiden Report, updated
version 2.0)

\bibitem[Karachentsev et al.(2000)]{Kara2000} Karachentsev, I. D.,
Sharina, M. E., \& Huchtmeier, W. K. 2000, A\&A, 362, 544

\bibitem[Kuntz et al.(2003)]{Kun03} Kuntz, K. D., Snowden, S. L.,
Pence, W. D., \& Mukai, K. 2003, ApJ, 588, 264

\bibitem[Larsen(1999)]{Lar99} Larsen, S. 1999, A\&ASup, 139, 393

\bibitem[Liedahl, Osterheld, \& Goldstein(1995)]{Lied95} Liedahl,
D.A., Osterheld, A.L., \& Goldstein, W.H. 1995, ApJL, 438, 115

\bibitem[Matonick \& Fesen(1997)]{MF97} Matonick, D. \& Fesen,
R. A. 1997, ApJS, 112, 49

\bibitem[McKee \& Ostriker(1977)]{MO77} McKee, C. \& Ostriker,
J. 1977, ApJ, 218, 148

\bibitem[Mewe, Gronenschild, \& van den Oord(1985)]{Mewe85} Mewe, R.,
Gronenschild, E.H.B.M., \& van den Oord, G.H.J. 1985, A\&AS, 62, 197

\bibitem[Mewe, Lemen, \& van den Oord(1986)]{Mewe86} Mewe, R., Lemen,
J.R., and van den Oord, G.H.J. 1986, A\&AS, 65, 511

\bibitem[Norman \& Ikeuchi(1989)]{NI89} Norman, C. \& Ikeuchi,
S. 1989, ApJ, 345, 372

\bibitem[Nulsen, Stewart, \& Fabian(1984)]{NSF84} Nulsen, P. E. J., Stewart,
G. C., \& Fabian, A. C. 1984, MNRAS, 208, 185

\bibitem[Ott et al.(2001)]{Ott01} Ott, J. et al. 2001, AJ, 122, 3070

\bibitem[Page et al.(2003)]{Page03} Page, M. J., Breeveld, A. A.,
Soria, R., Wu, K., Branduardi-Raymont, G., Mason, K. O., Starling,
R. L. C., \& Zane, S. 2003, A\&A, in press (astro-ph/0301027)

\bibitem[Roberts \& Warwick(2001)]{RobWar01} Roberts, T. P. \&
Warwick, R. S. 2001, in X-ray astronomy : stellar endpoints, AGN, and
the diffuse X-ray background, eds. N. E. White, G. Malaguti, \&
G. G.C. Palumbo (Melville, NY: American Institute of Physics), 474

\bibitem[Schlegel, Finkbeiner, \& Davis(1998)]{SFD} Schlegel, D. J.,
Finkbeiner, D. P., \& Davis, M. 1998, ApJ, 500, 525

\bibitem[Schlegel(1994)]{Sch94} Schlegel, E. M. 1994, ApJ, 434, 523

\bibitem[Schlegel, Blair, \& Fesen(2000)]{SBF00} Schlegel, E. M.,
Blair, W., P., \& Fesen, R. A. 2000, AJ, 120, 791

\bibitem[Summers et al.(2003)]{Summ03} Summers, L. K., Stevens, I. R.,
Strickland, D. K., \& Heckman, T. M. 2003, MNRAS, in press
(astro-ph/0303251)

\bibitem[Sutherland \& Dopita(1993)]{SuthDop93} Sutherland, R. \&
Dopita, M. 1993, ApJS, 88, 253

\bibitem[Townsley et al.(2002)]{Town02} Townsley, L. K.; Broos, P. S.;
Nousek, J. A.; Garmire, G. P., 2002, Nucl. Instr. \& Meth
Phys. Res. Sect. A, 486, 751.  

\bibitem[Tully(1988)]{Tul88} Tully, R. B. 1988, Nearby Galaxies
Catalog (Cambridge: Cambridge University Press)

\bibitem[Vedder et al.(1986)]{Ved86} Vedder, P. W., Canizares, C. R.,
Markert, T. H., \& Pradhan, A. K. 1986, ApJ, 307, 269

\bibitem[Walsh et al.(2002)]{Walsh02} Walsh, W., Beck, R., Thuma, G.,
Weiss, A., Wielebinski, R., \& Dumke, M. 2002, A\&A, 388, 7

\bibitem[Wang et al.(1995)]{Wang95} Wang, Q. D., Walterbos, R. A. M.,
Steakley, M. F., Norman, C. A., \& Braun, R. 1995, ApJ, 439 176

\bibitem[Young et al.(1989)]{Young89} Young, J. S., Xie, S., Kenney,
J. D. P.,\& Rice, W. L. 1989, ApJS, 70, 699

\bibitem[Young \& Scoville(1982)]{YS82} Young, J. \& Scoville,
N. 1982, ApJ, 258, 467

\end{thebibliography}
\end{document}